\documentclass[12pt]{iopart}   % Physical Biology
\usepackage{amssymb}
\usepackage{graphicx}
\usepackage{dcolumn}
\usepackage{bm}
\usepackage{booktabs}
\usepackage{ctable}
\usepackage{color}
\usepackage{multicol}
\usepackage{lineno}
\begin{document}
\title[Local activation of pC-collagen at individual fibrils]{Uniform spatial distribution of collagen fibril radii within tendon implies local activation of pC-collagen at individual fibrils}
\author{Andrew D Rutenberg$^1$, Aidan I Brown$^{1,2}$, Laurent Kreplak$^1$}

\address{$^1$ Department of Physics and Atmospheric Science, Dalhousie University, Halifax, NS, Canada, B3H 4R2}
\address{$^2$ Department of Physics, Simon Fraser University, Burnaby, BC, Canada, V5A 1S6}
\ead{\mailto{adr@dal.ca}}

\begin{abstract}
Collagen fibril cross-sectional radii show no systematic variation between the interior and the periphery of fibril bundles, indicating an effectively constant rate of collagen incorporation into fibrils throughout the bundle.  Such spatially homogeneous incorporation constrains the extracellular diffusion of collagen precursors from sources at the bundle boundary to sinks at the growing fibrils. With a coarse-grained diffusion equation we determine stringent bounds, using parameters extracted from published experimental measurements of tendon development.  From the lack of new fibril formation after birth, we further require that the concentration of diffusing precursors stays below the critical concentration for fibril nucleation. We find that the combination of the diffusive bound, which requires larger concentrations to ensure homogeneous fibril radii, and lack of nucleation, which requires lower concentrations, is only marginally consistent with fully-processed collagen using conservative bounds. More realistic bounds may leave no consistent concentrations.  Therefore, we propose that unprocessed pC-collagen diffuses from the bundle periphery followed by local C-proteinase activity and subsequent collagen incorporation at each fibril. We suggest that C-proteinase is localized within bundles, at fibril surfaces, during radial fibrillar growth. The much greater critical concentration of pC-collagen, as compared to fully-processed collagen, then provides broad consistency between homogeneous fibril radii and the lack of fibril nucleation during fibril growth. 
\end{abstract}

\noindent{\it Keywords\/}: collagen, fibril biogenesis, extracellular matrix, diffusion, C-proteinase
\maketitle
%\PACS 87.15.Vv \sep 87.16.Ka \sep 87.14.em
% 87.15.Vv	diffusion of biomolecules
% 87.16.Ka	filaments in subcellular structure
% 87.14.em	collagen biomolecules
%\linenumbers
%\today

%\pagebreak
%%%%%%%%%
\section{Introduction}%%%%%%%%
Mammalian tendons connect muscles to bones, are essential for locomotion, and can last a lifetime. Tendons contain hierarchical assemblies of collagen fibrils, with the most abundant collagen protein being type-I collagen. The three main models of tendon development and growth are the chick calcaneus tendon \cite{Marturano2013},  tendon-like constructs from human tenocytes \cite{Herchenhan2015}, and the mouse-tail tendon \cite{Kalson2015}.

Most ultrastructural studies have been performed in the mouse model. In mice, neighbouring tenocytes form extracellular bundles of fibrils delineated by a plasma membrane. These fibrils are produced in the embryonic stage by specialized cellular compartments called fibripositors \cite{Canty2004}, which disappear after birth \cite{Kalson2015, Humphries2008}. After pre-natal increase in both number of fibrils per bundle and average fibril area, the post-natal number of fibrils per bundle remains approximately constant. 

Following birth, the average fibril area increases by a factor of ten over the next six weeks \cite{Kalson2015}. This growth appears to be fed by procollagen stored in large diameter cellular compartments near the plasma membrane \cite{Humphries2008}. After procollagen assembly, N-propeptides can be removed by proteinases to produce pC-collagen (pC) or C-propeptides can be removed to form pN-collagen (pN). Fully-processed collagen has both N- and C-propeptides removed. We refer to any of these forms of collagen as a potential precursor for fibril growth. While intracellular collagen appears to be largely  pC \cite{Humphries2008}, the precursor that is secreted into the extracellular space to facilitate fibril growth has not been directly characterized in either form or abundance. 

Strikingly, the spatial distribution of fibril radii remains uniform from the earliest embryonic stages until adulthood--- with no visible differences in fibril radii between the interior and the periphery of fibril bundles. This cross-sectional spatial homogeneity is apparent in tail tendons of mice \cite{Kalson2015}, and in cross-sections of the anterior cruciate ligament (ACL) of humans \cite{Suzuki2015}. We expect that ongoing radial growth of fibrils  \cite{Kalson2015, PattersonKane1997, Goh2012} is facilitated by diffusive transport of collagen precursors from the bundle periphery, since there is no mechanism for either active transport or precursor synthesis within extracellular fibril bundles. Given diffusive transport, then the observed homogeneity of fibril radii should place biophysical constraints on the precursor abundance. Precursor abundance must be sufficiently large to support homogeneous radial growth of fibrils. Independently, keeping precursor abundance {\em below} the critical concentration of nucleation is essential to prevent unchecked fibril formation. We explore these constraints in this paper.

%%%%%%%%%
%%%%%%%%%
\section{Results}

%%%%%%%%%%%%
\subsection{Diffusive constraint}
The fibril bundles have homogeneous fibril radii throughout their cross-section. We model the diffusive environment using a coarse-grained effective medium with a precursor diffusivity $D_{col}$. We also assume homogeneity of the coarse-grained ``sinks'' of precursor transport --- i.e. the local growth of fibrils. This coarse-graining is appropriate to model transport at length-scales larger than typical fibril separations. The reaction-diffusion equation for collagen precursors is then 
\begin{equation}
	\partial{\rho}/\partial t = D_{col} \nabla^2 \rho - s,
		\label{eq:diffusion}
\end{equation}
 where $\rho(\vec{r},t)$ is the cross-sectional number density of precursors,  $s$ is the spatially homogeneous rate (per unit area) of precursor incorporation onto fibrils, and $\nabla^2$ is a Laplacian.   Since fibril growth takes weeks, we take a quasi-static approach in which we assume the precursor density field remains in steady-state as the bundle size slowly changes. We take cylindrical bundles of radius $R$, and impose rotational symmetry, so that we can solve Eqn.~\ref{eq:diffusion} in cylindrical coordinates with a fixed precursor density at the bundle boundary $\rho(R)=C$.  For steady-state, the radial equation is $0 = D_{col}( \partial^2 \rho/\partial r^2+ (\partial \rho / \partial r)/r)-s$. The solution is
 \begin{equation}
 	\rho(r) = s (r^2-R^2)/(4D_{col}) + C
	\label{eq:solution}
\end{equation}

If $\Delta\rho$ is the change in concentration from the edge of the fibril bundle to the centre of the fibril bundle, then a measure of heterogeneity is $\Delta\rho/\rho(R) = 1 - \rho(0)/\rho(R)$. This will be maximized, with $\Delta\rho/\rho(R)=1$, when the concentration at the centre of the bundle vanishes, and minimized, with $\Delta\rho/\rho(R)=0$, when the concentration at the centre of the bundle and at the edge of the bundle are equal. For homogeneous fibril radii, the measure of heterogeneity must be kept low. If we insist that $\Delta \rho/\rho(R) \leq f$, for some acceptable heterogeneity factor $f \in (0,1]$, then using Eq.~\ref{eq:solution} we have
\begin{equation}
	D_{col} C \geq S / (4 \pi f) , 
	\label{eq:bound}
\end{equation}
where $S \equiv s \pi R^2$ is the total rate of precursor consumption - and therefore the rate of their introduction at the edge of the bundle - needed to support growth of all of the fibrils in a bundle. This provides a combined bound on the number density and diffusivity of collagen precursors.

%%%%%%%%%%%%
\begin{figure}[h!] 
\centering
  \hspace{-0.3in}
  \begin{tabular}{c}
 \vspace{0.00in} \hspace{-0.0in} \includegraphics[width=3.0in]{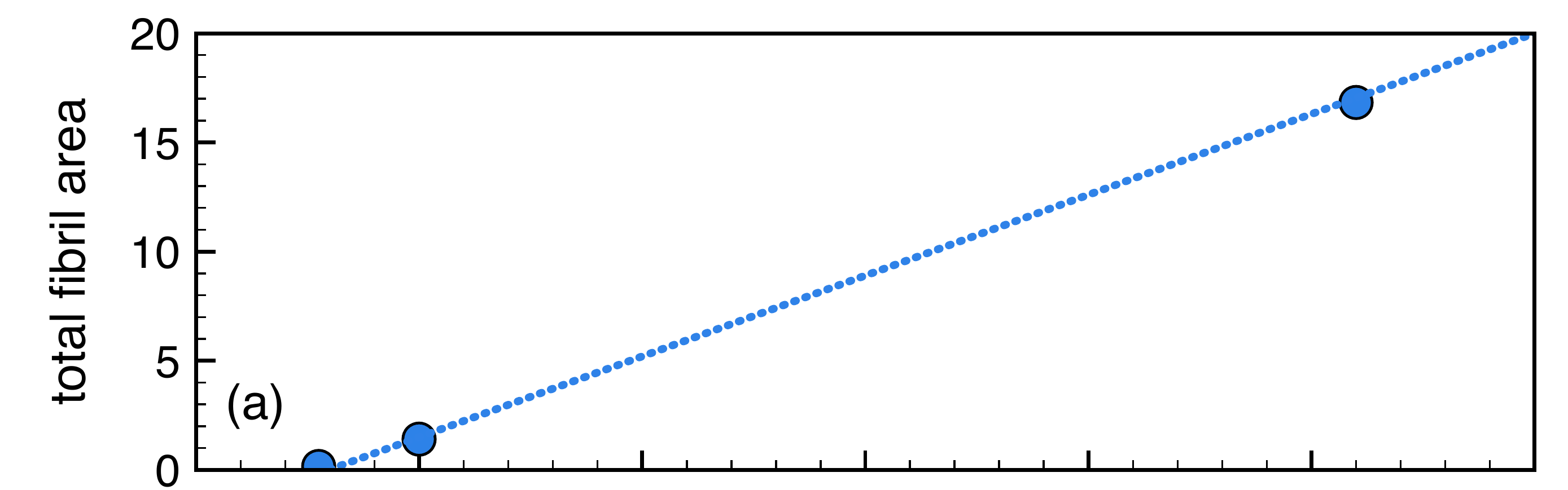} \\
 \vspace{-0.00in} \hspace{-0.0in} \includegraphics[width=3.0in]{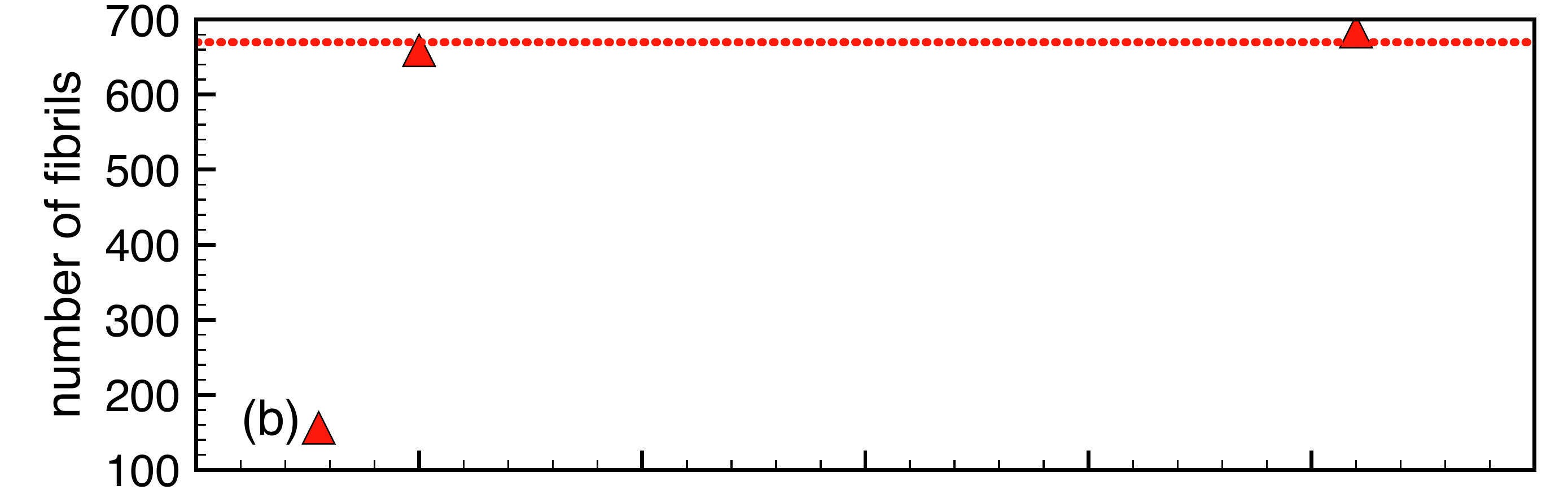} \\
 \vspace{-0.00in} \hspace{-0.0in} \includegraphics[width=3.0in]{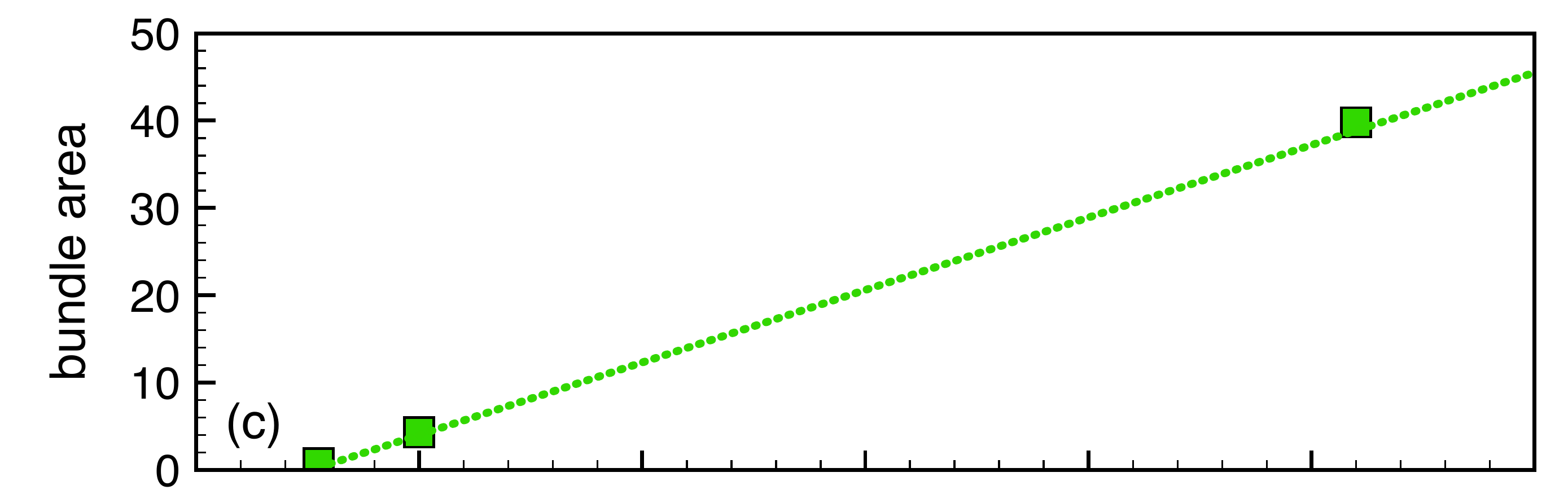} \\
 \vspace{-0.00in} \hspace{-0.0in} \includegraphics[width=3.0in]{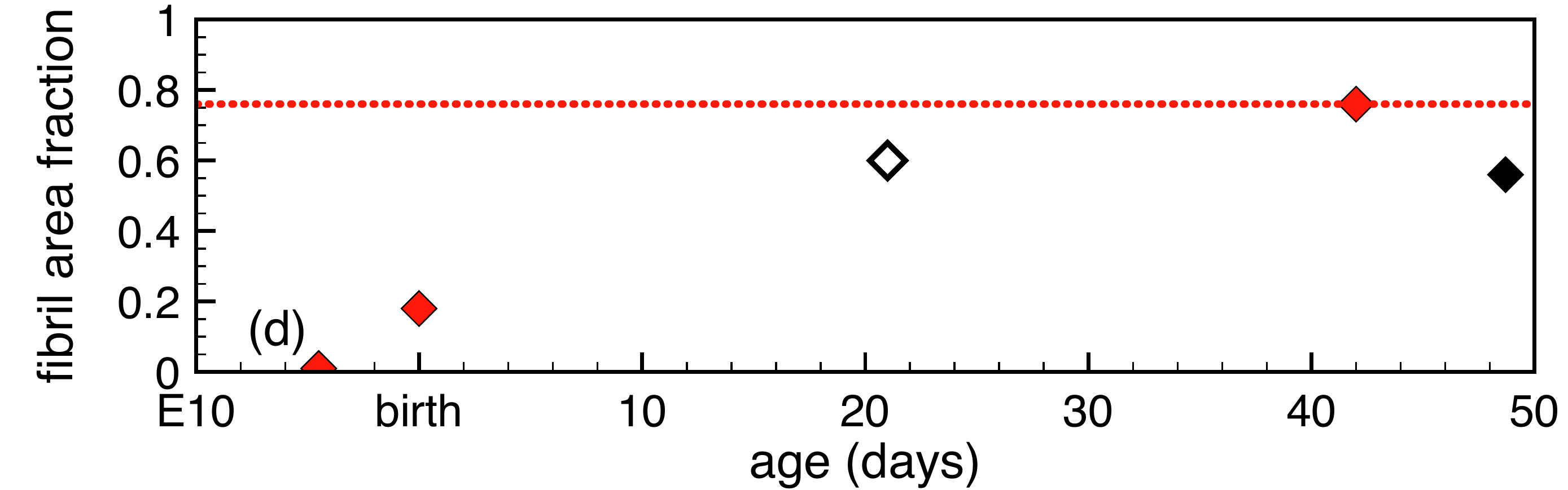}
\end{tabular}
\caption{\label{fig:figure1}  Collagen fibril growth parameters. (a) Total fibril area vs.\ age (blue circles, in $\mu$m$^2$, from \cite{Kalson2015}. Dotted blue line indicates a growth rate of $dA/dt \approx 0.37 \mu$m$^2$/day, as used in the text. (b) Number of fibrils per bundle (red triangles, from \cite{Kalson2015}. Dotted red line indicates an approximately constant number of fibrils after birth, as used in the text. (c) Average bundle area (green squares, in $\mu$m$^2$, from \cite{Kalson2015}. Dotted green line indicates a growth rate of $d(\pi R^2)/dt \approx 0.83 \mu$m$^2/$day, as used in the text. (d) Fibril area fraction (red diamonds from \cite{Kalson2015}, empty diamond from \cite{Robinson2004}, black diamond from \cite{Goh2008}. Dotted red line indicates the maximal area fraction $\phi_{ob} \approx 0.76$, as used in the text.}
\end{figure}
   
%%%%%%%%%%%%%%%%%%%%
\subsection{Precursors needed for growth, $S$} %%%%%%%%%%
The number rate of precursor consumption necessary for growth, $S$, is determined by the area growth rate of fibrils:
\begin{equation}
	S =  (d A/dt) /\sigma,
\end{equation}
where $d A/dt$ is the net growth of fibril area for all fibrils in a bundle following fibril formation, and $\sigma$ is the area per precursor. The cross-sectional diameter of one collagen molecule is approximately $1.08$ nm \cite{Hulmes1995}, so that $\sigma = 0.916$ nm$^2$. 

From Fig.~\ref{fig:figure1}, after birth (at $t=20$ days) the number of fibrils per bundle is approximately fixed and so growth of total fibril area in a bundle is due to the increasing size of individual fibrils. We approximate the growth of the total fibril area as linear in time, with $d A/dt \approx 0.37$ $\mu$m$^2$/day as indicated by the dashed line in Fig.~\ref{fig:figure1} (a). Linear growth provides a (conservative) lower bound on the maximal growth rate and hence on S. This fibril area growth rate gives $S \approx 4.7$/s, as the number of fully-processed collagen molecules needed per second to support growth of each individual bundle.

%%%%%%%%%%%%%%%
\subsection{Collagen diffusivity, $D_{col}$} %%%%%%%%%%%%%%%%%%%%%%
The self-diffusion of suspensions of colloidal rod-like molecules such as collagen \cite{vanbruggen1997, vanbruggen1998} depends on particle shape, via the ratio of length to diameter $p \equiv \ell/d$, and on volume fraction $\phi_{vol}$. In the dilute limit with long rod-like molecules, with   $\phi_{vol} \approx 0$ and $p \gg 1$, we have \cite{vanbruggen1997, vanbruggen1998}
\begin{equation}
	D_0 = \frac{k_B T}{3 \pi \eta \ell} \left[ \ln{p} + 0.316 + 0.5825/p+0.050/p^2 \right],
	\label{eq:D}
\end{equation}
where $\eta$ is the dynamic viscosity of the solvent fluid, and $k_B T$ is the characteristic thermal energy-scale. Essentially, this is Stokes-Einstein thermal diffusion  with the characteristic particle size given by the particle length $\ell$, and with drag due to end-effects included.  While the effects of concentrated solutions of elongated particles are not as precisely characterized, increasing $\phi_{vol}$ always leads to lower diffusivity. For example, when $ p \phi_{vol} \approx 1$, the diffusivity decreases approximately ten-fold \cite{vanbruggen1997, vanbruggen1998}.
 
For diffusion around obstructions, such as provided by collagen fibrils, an additional obstruction or tortuosity factor, $A_\perp$, further reduces the diffusivity so that $D = D_0 \times A_\perp$.  For cylindrical obstructions in a hexagonal lattice, this factor has been calculated and verified with detailed simulations \cite{Johannesson1996}:
\begin{equation}
	A_\perp = \frac{1}{1-\phi_{ob}} \left( 1 - \frac{ 2\phi_{ob}}{1 + \phi_{ob} - 0.07542 \phi_{ob}^6} \right),
	\label{eq:obstruct}
\end{equation}
where $\phi_{ob}$ is the area fraction of obstructions. $A_\perp$ monotonically decreases with increasing $\phi_{ob}$. The fibril area fraction increases with age so we conservatively use the value at age 6w, $\phi_{ob} \approx 0.76$ \cite{Kalson2015} so that $A_\perp \approx 0.54$. 

For collagen, with $\ell \approx 300$ nm and $d \approx 1.08$ nm, we have $p \approx 278$. Assuming a dilute suspension of collagen in water at body temperature, with  $\eta_{water} \approx 0.67$ cP, $k_B T \approx 4.3$ pN nm, and $A_\perp =0.54$,  we combine equations \ref{eq:D} and \ref{eq:obstruct} to obtain
\begin{equation}
	D_{col} = D_0 \times A_\perp \approx 7.3 \mu{\rm m}^2/{\rm s}.
\end{equation}
This precursor diffusivity will be significantly reduced when volume fractions reach $p \phi_{vol} \approx 1$, and with larger precursors.   Other solutes can also significantly thicken the solvent, for example cellular cytoplasm is approximately five times as viscous as water \cite{Verkman2002}. Because of these possible effects, our value of $D_{col}$ is a conservative over-estimate of the extracellular precursor diffusivity.

We can now check the quasi-static approximation underlying Eqn.~\ref{eq:diffusion}. For it to be valid we need the neglected term ($\partial \rho/\partial t$) to be much smaller in magnitude than the incorporation rate $s$. Using the solution Eqn.~\ref{eq:solution} to evaluate $\partial \rho/\partial t$, the ratio of the terms is 
\begin{equation}
		\left| \partial \rho/\partial t \right|/s = d(\pi R^2)/dt/(4 \pi D_{col}) \approx 1.0 \times 10^{-7},
\end{equation}
where $d(\pi R^2)/dt \approx 0.83 \mu$m$^2/{\rm day}$ is the rate of bundle area increase from birth to 6w \cite{Kalson2015}. We see that the dimensionless ratio is vanishingly small, i.e. the quasi-static approximation was well justified. Fibrils and bundles grow slowly compared to diffusion timescales.  

%%%%%%%%%%%%%%%%%
\subsection{Conservative lower-limit of precursor abundance} %%%%%%%%%%%%%%%%%%%%%%
For now, we set the heterogeneity factor $f=1$, which corresponds to a concentration of nearly zero at the centre of the bundle. This provides the most conservative (lowest) bound on the precursor concentration $C$ from Eqn.~\ref{eq:bound}. Using $D_{col}$ we then obtain $C \gtrsim 0.051/\mu$m$^2$. For precursors aligned with the fibril axis, the corresponding bulk number-density is $\rho_{lower} = C/\ell \approx 0.17/\mu$m$^3$. $\rho_{lower}$ is the minimum concentration required at the edge of fibril bundles to support spatially uniform growth. Using $\ell$ and $d$ for collagen, we estimate $p \phi_{vol} \approx 1 \times 10^{-5}$ at this lower bound, which is comfortably in the dilute limit assumed in Eqn.~\ref{eq:D}.

Whole cell proteomics can estimate collagen abundance {\em in vivo}. PaxDB \cite{Wang2012, Wang2015} provides an online resource for protein abundances, and integrated values over all listed studies estimate the abundance of the $\alpha 2$ chain collagen I in human cell lines to be $39$ppm, which is comparable to the whole-body average of  $115$ppm (or $81$ppm for mice). We take the smaller cell-line abundance as a crude upper bound on non-fibrillar collagen abundance. By using the protein number density of approximately  $2.7 \times 10^6/ \mu $m$^3$ in mammalian cells \cite{Milo2013}, we then obtain a bulk number density of collagen of $\rho \approx 105 / \mu $m$^3$. We see that our estimate of $\rho_{lower}$ is two orders of magnitude less than, and so consistent with, the estimated bulk number densities from cell-line studies.

Using a molecular weight of collagen $M \approx 3.0 \times10^5$ (g/mole) \cite{Petruska1964}, and using the number density $\rho_{lower}=0.17/\mu$m$^2$, we obtain a conservative lower-limit estimate of the mass density of diffusing collagen of $\rho_{mass} = 0.08 \mu $g/ml.  This is  below the measured critical concentration for self-assembly of fully-processed collagen into fibrils, $0.42 \mu $g/ml at 37$^\circ$C \cite{Kadler1987}, and more than four orders of magnitude below the critical concentration for pC collagen, which is at least $0.5$ mg/ml \cite{Kadler1987}.

%%%%%%%%%%%%%
\subsection{Using the heterogeneity factor $f<1$}
For fully processed collagen, we expect diffusion-limited fibril growth in which  the radial fibril growth rate is proportional to the local precursor concentration. The heterogeneity factor then directly determines the heterogeneity of fibril radii between the bundle core and periphery. With a heterogeneity factor $f<1$ we have $\rho_{lower} \approx (0.17/f)$ $\mu$m$^{-3}$ and $\rho_{mass} =(0.08/f) \mu $g/ml. Given the apparent homogeneity of radii \cite{Kalson2015, Suzuki2015}, we would estimate $f \lesssim 0.2$. This puts us right at the critical concentration for fully-processed collagen, though still more than three orders of magnitude below the critical concentration for pC collagen.

We have also made conservative assumptions that --- if relaxed --- would further increase $\rho_{lower}$ and $\rho_{mass}$. We worked in the dilute limit for Eqn.~\ref{eq:D} and took the solvent viscosity to be that of water.  We assumed that the precursor is a single collagen molecule, since larger precursors would diffuse even slower. Furthermore, fluctuations in bundle radius or precursor concentration could lead to fibril nucleation near the nucleation threshold, and so argues for higher bounds.  Therefore, although we have not entirely ruled out diffusing collagen as supplying fibril growth, our calculations indicate that fully processed collagen is not a good candidate for the diffusing precursor driving extracellular fibril growth after birth.  In contrast, pC collagen is a viable candidate.

%%%%%%%%%%%%%
\section{Discussion} 

Procollagen is first processed intracellularly by N-proteinase in post-Golgi compartments \cite{CantyLaird2012} to produce pC-collagen, which still has its COOH-terminal propeptide.  We have shown that extracellular diffusion of pC-collagen from bundle periphery to core could support spatially uniform fibril growth while remaining orders of magnitude below its critical concentration. In contrast, for diffusion of fully processed collagen it would be at or above its critical concentration. Given the lack of fibrillogenesis after birth \cite{Kalson2015}, and the general observation that fibrillogenesis is not observed in the absence of collagen binding partners \cite{Kadler2008}, we therefore propose that the diffusive flux from bundle peripheries to growing fibrils is dominated by pC-collagen. 

We further propose that pC-collagen is processed at the growing fibrils. This would imply that fibril growth could be rate-limited by C-proteinase activity, with $f=1$. Indeed,  {\em in vivo} fibril growth studies from the spine ligament of sea urchin point towards a surface mediated growth process that appears rate-limited \cite{Trotter2000}. However, the enormous critical concentration for pC compared to fully-processed collagen allows it to satisfy our bounds at even very small values of the heterogeneity factor $f$. 

%%%%%%%%%%%%
\begin{figure}[h!] % figure 2
\centering
  \hspace{-0.3in}
  \begin{tabular}{c}
  \vspace{-1.00in} \hspace{-0.0in} \includegraphics[width=3.0in]{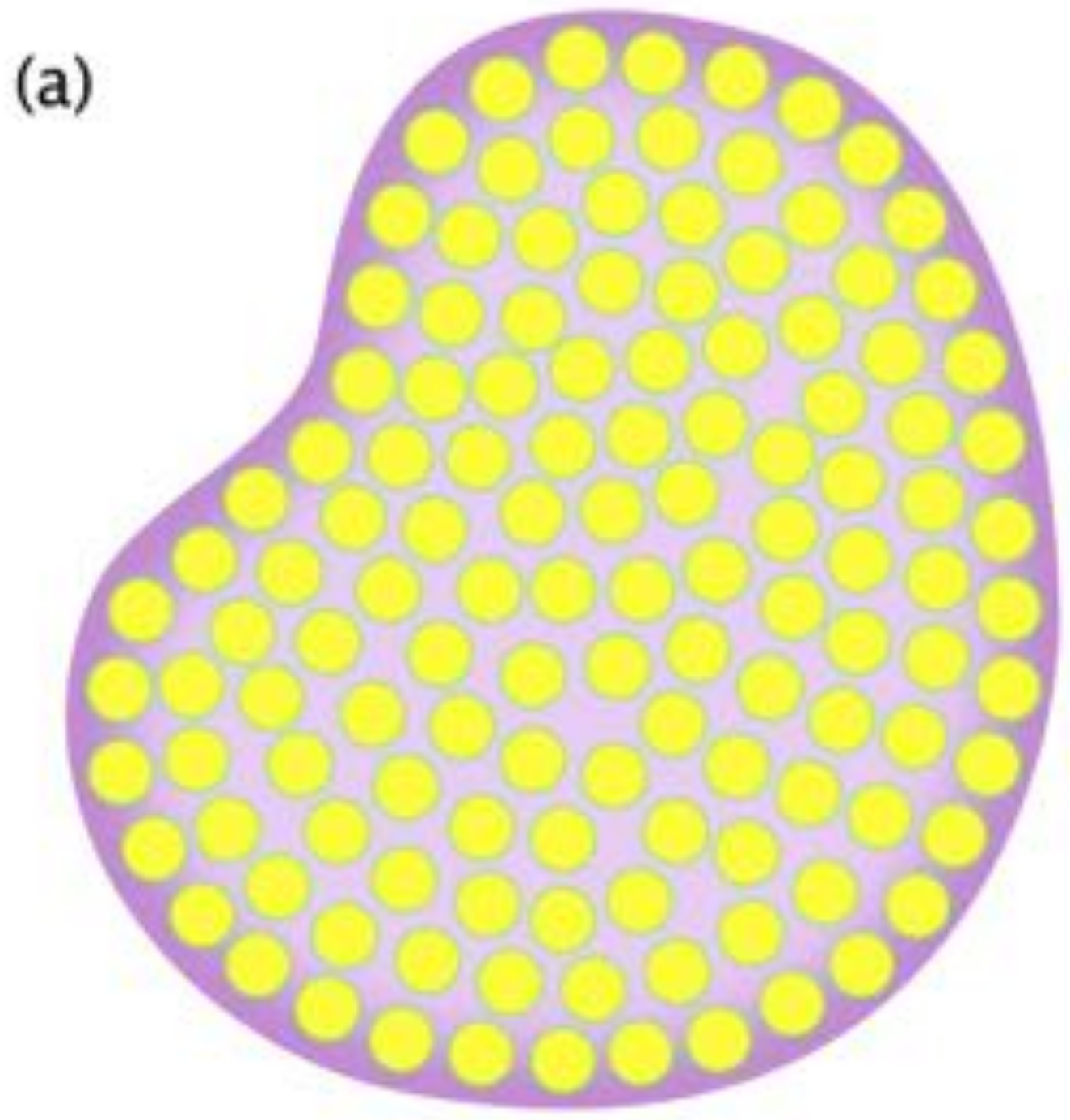} \\
 \vspace{-.50in} \hspace{-0.0in} \includegraphics[width=3.0in]{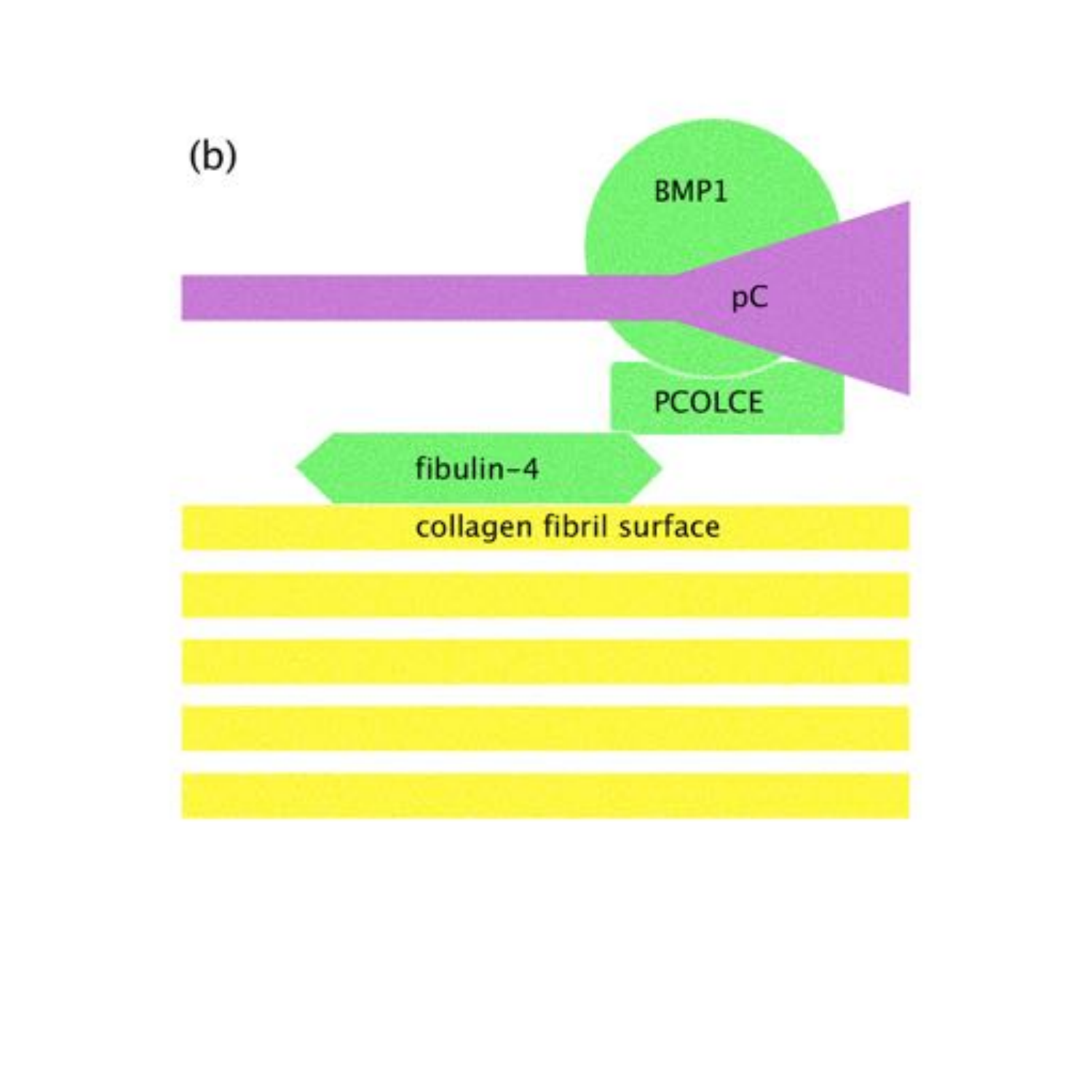} 
\end{tabular}
\caption{\label{fig:figure2}  Proposed model of collagen fibril growth. (a) A cross-section through a bundle of fibrils (yellow) highlights the extracellular concentration gradient of precursor collagen pC (purple) that decreases from the bundle exterior to  interior.  The conversion of pC into assembly competent collagen happens inside the bundle, at the surface of each fibril (green ring). (b) A lateral view near the surface of a collagen fibril (yellow bars). Mechanistically, we propose that fibulin-4 (green signpost shape) binds to the surface of  collagen fibril, where it forms a complex with a protein enhancer of C-proteinase activity (PCOLCE,  green rectangle), a C-proteinase (BMP-1, green circle), and pC-collagen (purple). This (green) complex locally catalyzes the cleavage of the C-propeptide leaving a collagen molecule that can then incorporate into the fibril surface.}
\end{figure}

Currently, the evidence for C-proteinase activity at growing fibrils is indirect but consistent with our proposals. Knocking out the PCOLCE1 enhancer of procollagen C-proteinase in mice leads to fibrils with distinctively knobbly appearance \cite{Steiglitz2006}. If C-proteinase activity took place only far from fibrils, then reduced activity would lead to slower growth but not to irregular cross-sections. Hence, the irregularity of the fibril shape hints at procollagen C-proteinase activity occurring in close proximity to each fibril surface.   Irregular fibril cross-sections are also observed in people with mutations in the {\em BMP1} gene encoding a human C-proteinase \cite{Syx2015}. What has been missing is a protein that links pC-collagen and BMP1 or PCOLCE1 to the surface of growing fibrils. A recent knockout study of fibulin-4, a protein essential to the formation of elastic fibres in the aorta, also shows an irregular fibril cross-section phenotype \cite{Papke2015}. Strikingly, their model of fibulin-4 shows colocalization of pC-collagen, fibulin-4, BMP1, and PCOLCE1 at the fibril surface -- with fibulin-4 acting as a scaffold protein \cite{Papke2015}. 

In summary, we propose that, after birth, the growth of collagen fibrils within tendons is supported by diffusing pC-collagen secreted by cells lining the sides of collagen bundles. We further propose that activation of the pC-collagen is controlled by C-proteinase activity at fibril surfaces. These two  proposals, illustrated in Fig.~2, would satisfy the observation of spatial uniformity of fibril radius from bundle core to periphery, together with the requirement that diffusing precursors remain below their critical concentration.

Our analysis and proposals highlight the significance of the spatially uniform distribution of fibril radii in constraining the nature of the diffusing precursor for collagen fibril growth. Better experimental characterization of the heterogeneity of fibril radii, and hence of $f$, would tighten the bounds we have discussed here --- as would direct experimental determination of the extracellular concentration of precursors. The current evidence, and our analysis, points towards C-proteinase activity at the fibril surface -- so it would also be desirable to assess any significant localization of C-proteinases outside of fibril bundles. 

Our picture of the fibril radial growth process within tendon ignores many regulators of collagen assembly \cite{Kadler2008}, though it is not always easy to separate the roles of fibrillogenesis with fibril growth, which is our focus here.  For our picture, regulators which affect procollagen processing, diffusion $D$, the critical concentration of diffusing precursors, or the local growth mechanism $f$, are important --- as are any that phenomenologically affect the observed spatial homogeneity of fibril radii. Of particular note are proteoglycans such as decorin and biglycan that regulate fibril radius in vivo \cite{Danielson1997, Corsi2002}, and extracellular glycoproteins such as COMP that can enhance fibril formation in vitro \cite{Halasz2007}. The proteoglycans are directly bound to the fibril surface so could compete for binding sites with the BMP1/PCOLCE1/fibulin-4 complex shown in Fig.2. This could affect, for example, the fibril growth rate. COMP enhances fibril assembly in vitro by binding to the soluble precursors rather than the growing fibrils \cite{Halasz2007}, and this could affect the hydrodynamic size of the diffusing collagen precursor and hence its diffusivity $D$.  Our physically based model provides a clear foundation for thinking about the mechanisms for changes of extracellular fibril growth, as different collagen regulators are perturbed.

This work was supported by the Natural Sciences and Engineering Research Council of Canada (NSERC) with operating grant numbers RGPIN-2014-06245 (to ADR) and RGPIN-355291-2013 (to LK). AIB thanks NSERC, the Sumner Foundation, and the Killam Trusts for fellowship support.

%%%%%%%%%%%%%%%%%%
\section*{References}
 \bibliographystyle{unsrt} %.bst file
  \bibliography{collagendiffusion}  
\end{document}